\documentclass[journal=ancac3,manuscript=article]{achemso} 

\usepackage{amsmath}
\usepackage{graphicx}
\usepackage{siunitx} \DeclareSIUnit{\litre}{l}
\usepackage[version=4]{mhchem}
\usepackage{textgreek}

\usepackage[dvipsnames]{xcolor} \usepackage{soul}

\author{Michael Foltýn}
\affiliation[CEITEC]
{Brno University of Technology, Central European Institute of Technology, Purky\v{n}ova 123, 612 00, Brno, Czech Republic}
\alsoaffiliation[TFS]
{Thermo Fisher Scientific, Vlastimila Pecha 12, 627 00  Brno, Czech Republic}

\author{Michal Kvapil}
\affiliation[CEITEC]
{Brno University of Technology, Central European Institute of Technology, Purky\v{n}ova 123, 612 00, Brno, Czech Republic}
\alsoaffiliation[UFI]
{Brno University of Technology, Faculty of Mechanical Engineering, Institute of Physical Engineering, Technická 2, 616 69, Brno, Czech Republic}

\author{Kristýna Bukvišová}
\affiliation[TFS]
{Thermo Fisher Scientific, Vlastimila Pecha 12, 627 00  Brno, Czech Republic}

\author{Tomáš Šikola}
\affiliation[CEITEC]
{Brno University of Technology, Central European Institute of Technology, Purky\v{n}ova 123, 612 00, Brno, Czech Republic}
\alsoaffiliation[UFI]
{Brno University of Technology, Faculty of Mechanical Engineering, Institute of Physical Engineering, Technická 2, 616 69, Brno, Czech Republic}

\author{Michal Horák}
\affiliation[CEITEC]
{Brno University of Technology, Central European Institute of Technology, Purky\v{n}ova 123, 612 00, Brno, Czech Republic}
\email{michal.horak2@ceitec.vutbr.cz}

\title{Lead nanoparticles, the deep-ultraviolet to near-infrared plasmonic platform}

\begin{document}


\begin{abstract}

Among the other non-noble metals, lead (Pb) is a material of particular interest for plasmonic applications in the deep ultraviolet spectral region. However, experimental studies on its plasmonic performance have not yet been conducted. In this work, the dependence of the optical properties of spherical lead nanoparticles on their diameter is demonstrated. The plasmonic performance of chemically synthesized lead nanoparticles is evaluated at the single-particle level by means of a combination of scanning transmission electron microscopy and electron energy loss spectroscopy. Our findings demonstrate that these nanoparticles support localized surface plasmon resonances across the entire spectrum from near-infrared to deep-ultraviolet. This range was identified as the most extensive among all plasmonic elemental metals, extending to wavelengths below 200 nanometers. Consequently, lead nanoparticles exhibit stable plasmonic performance over a remarkably broad wavelength range, thereby substantiating their potential as a multispectral plasmonic platform.

\end{abstract}


\section{Introduction}

Collective oscillations of free electrons in metallic nanostructures, known as localized surface plasmon resonances (LSPRs) \cite{10.1021/jp026731y}, are commonly used in a wide field of applications, including microscopy \cite{10.1038/s41467-019-13230-1}, catalysis \cite{10.1021/jacs.3c14586}, biosensing \cite{10.1021/acs.nanolett.5b05316}, 3D real-time sensing \cite{10.1021/acsnano.9b09508}, metamaterials \cite{10.1016/j.cej.2025.168812}, and quantum computing \cite{10.1038/s41534-019-0150-2}. Nevertheless, a considerable number of these applications are limited to the near-infrared and visible electromagnetic spectrum. Despite the numerous attractive applications, their use in the ultraviolet region is limited by interband transitions and the plasmon damping typical in traditional plasmonic metals such as gold \cite{10.1021/nn102166t}. When considering the ultraviolet region, alternative metals (including non-noble metals \cite{10.1039/d6ra02381a}), such as aluminum \cite{10.1021/ja210446w, 10.1021/am505511v, 10.1038/srep19887, 10.1038/s41598-021-84550-w}, rhodium \cite{10.1021/nl5040623, 10.1039/d4nh00449c}, gallium \cite{10.1021/nn5072254, 10.1021/acs.jpclett.3c00094, 10.1021/acs.jpclett.5c02035}, magnesium \cite{10.1039/c6cc06800f, 10.1021/acs.nanolett.3c03219}, and tin \cite{10.1021/acs.jpcc.8b10851} have been identified as suitable materials for UV plasmonic applications.

Another promising plasmonic material for the ultraviolet spectral region is lead. With its relatively low imaginary part of the dielectric function and negative real part of the dielectric function extending to deep ultraviolet wavelengths \cite{10.1039/c3cp43856b}, lead could support LSPRs even at very short wavelengths. A comparison of experimental dielectric functions of lead from the literature, namely that of Werner et al. \cite{10.1063/1.3243762}, Mathewson \& Myers \cite{10.1088/0031-8949/4/6/009}, Ordal et al. \cite{10.1364/ao.26.000744}, Golovashkin \& Motulevich \cite{Golovashkin1968}, and Lemonnier et al. \cite{10.1103/physrevb.8.5452}, is shown in Figure~\ref{FigS1}. The main limiting factor is their energy range. The ultraviolet spectral region is covered by the dielectric functions of Werner et al. \cite{10.1063/1.3243762} and Lemonnier et al. \cite{10.1103/physrevb.8.5452} whose real part reaches values below $-2$ up to \SI{6}{\electronvolt} and \SI{7}{\electronvolt}, respectively, and is negative up to \SI{12}{\electronvolt}. However, it should be noted that the dielectric function of Werner et al. is distinct from the other four. Theoretical quality factors of LSPRs, defined as $Q_\mathrm{LSPR}=-\Re{(\epsilon)}/\Im{(\epsilon)}$, range from 3 to 7 in the visible and ultraviolet spectral range, while the others indicate that lead would be hardly promising plasmonic material with theoretical quality factors of LSPRs below 2.5 in the entire energy range. This suggests the possibility of a redefinition of the dielectric function of lead to facilitate exact theoretical assumptions.

Experimental studies on lead plasmonics are rather rare in the literature. To date, they have been limited to a single investigation of the plasmonic response of lead films deposited on nanostructured substrates, which demonstrated the potential for lead to be used as an alternative plasmonic material within the visible spectral range \cite{10.1063/5.0016131}. Plasmons were also investigated in very thin lead nanowires on a silicon substrate \cite{10.1103/physrevb.84.205402} and the synthesis of lead nanoparticles was thoroughly discussed \cite{10.1021/nn102009g, 10.1039/c5ra18054f}. However, no experimental studies have yet been performed on lead nanostructures as a standalone plasmonic platform, nor has there been a critical evaluation of lead as a plasmonic material across its entire spectral range.

In this work, we present an experimental analysis of the optical properties of chemically synthesized lead spherical nanoparticles at the single particle level using electron energy loss spectroscopy in a scanning transmission electron microscope (STEM-EELS). The experimental findings are further supported by numerical simulations, thereby enhancing their reliability and robustness. We have demonstrated the spectral tunability of LSPRs in individual spherical lead nanoparticles from the near-infrared to deep ultraviolet spectral region in correlation with the nanoparticle diameter. Our results show that lead exhibits the most extensive spectral tunability within the comprehensive array of plasmonic elemental metals that have been examined.


\section{Results and Discussion}

\begin{figure}[t]
    \centering
    \includegraphics[width=1\linewidth]{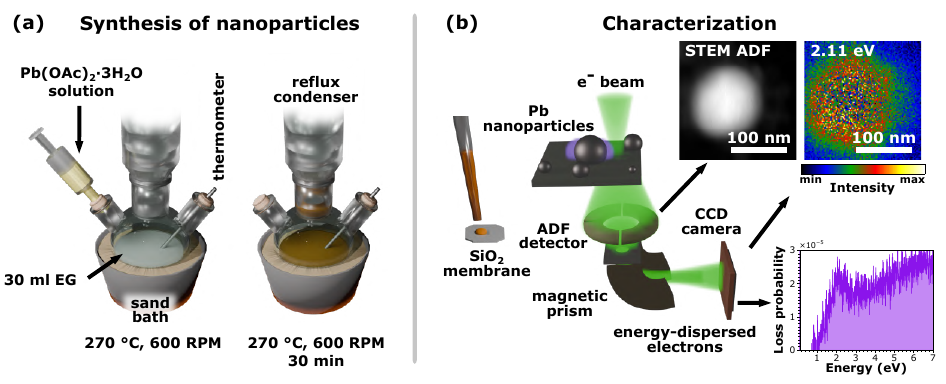}
    \caption{Schematic workflow of the synthesis and chracterization of lead nanoparticles. (a) The spherical lead nanoparticle solutions were prepared by injection of lead acetate trihydrate solution into preheated tetraethylene glycol. Additionally, trisodium citrate was injected as well, to obtain smaller average nanoparticles. The solution was then left to react for 30 minutes and cooled down to room temperatures. (b) The morphology of lead nanoparticles was obtained from annular dark-field (ADF) STEM micrographs, and their plasmonic resonances were inspected using STEM-EELS.}
    \label{Fig1}
\end{figure}

The synthesis of lead spherical nanoparticles was accomplished through the thermal decomposition of lead acetate trihydrate in tetraethylene glycol (TEG) at elevated temperatures. The schematic representation of the fabrication and characterization workflow is shown in Figure~\ref{Fig1}. The synthesis process involves the injection of lead acetate trihydrate, which has been dissolved in a minimal amount of TEG, into the preheated TEG at a temperature of \SI{270}{\celsius}. Trisodium citrate can be injected as an optional additional treatment. Trisodium citrate functions as a weak electrostatic surfactant \cite{10.1038/s41467-020-19164-3}, resulting in the formation of smaller lead nanoparticles. For further details, see Methods. The formation of lead nanoparticles is indicated by a relatively rapid color change to orange-brown and the formation of a black precipitate at the bottom of the vial. After 30 minutes, the heating is turned off and then the solution is left to gradually cool to room temperature. The prolonged cooling process enables the formation of larger nanoparticles, resulting in a more extensive distribution of the nanoparticle diameters. This facilitates the study of the optical properties of lead nanoparticles across a wide range of diameters. The solutions containing citrate are designated as the Pb-C solution, while the citrate-free solution is referred to as the Pb-NC solution. Both solutions were thoroughly washed and subsequently diluted in methanol. The washed solution was subsequently drop-casted onto commercially available \ce{SiO2} membranes for transmission electron microscopy (TEM) and plasma cleaned in an oxygen-argon atmosphere.

\begin{figure}[p]
    \centering
    \includegraphics[width=1\linewidth]{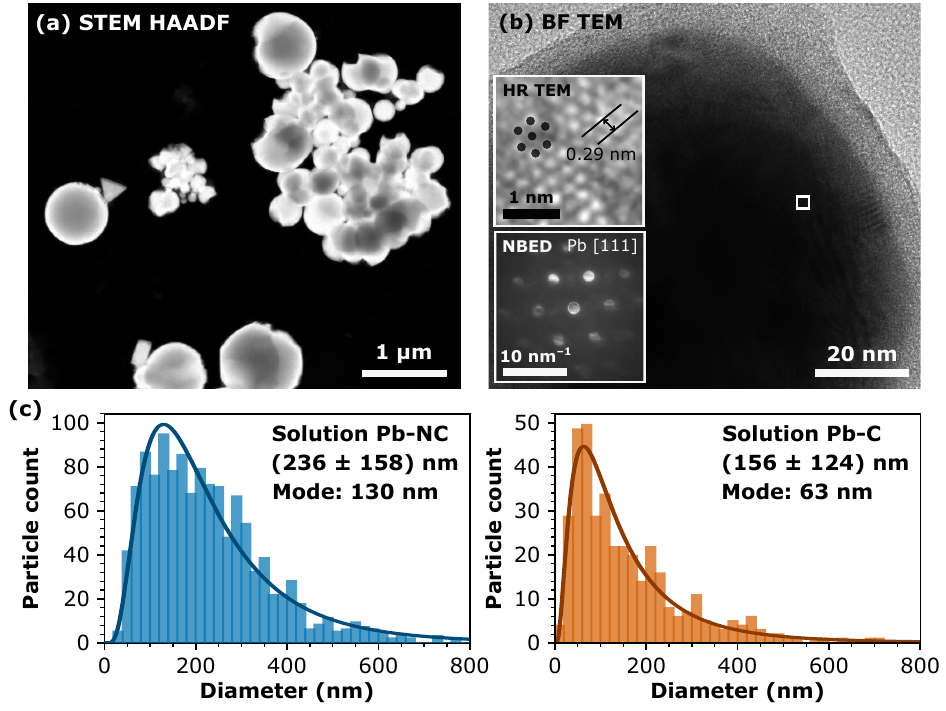}
    \caption{Morphology, crystallinity, and size distribution of synthesized lead nanoparticles. (a) HAADF STEM micrograph showing their typical shape and size distribution. (b) TEM micrograph of a \SI{100}{\nano\meter} nanoparticle. The inset shows Fourier-filtered image of the atomic columns from the area marked by the white rectangle. A NBED diffractogram from the same area indicates the crystal orientation of [111]. (c) Histograms with fitted log-normal curve showing the size distribution of lead nanoparticles in the Pb-NC and Pb-C solution. The Pb-NC solution contains nanoparticles with a mean diameter of \SI{236}{\nano\meter}, its standard deviation of \SI{158}{\nano\meter}, and its mode of \SI{130}{\nano\meter}. The size distribution of the Pb-C solution reads a mean diameter of \SI{156}{\nano\meter}, its standard deviation of \SI{124}{\nano\meter} and its mode of \SI{63}{\nano\meter}.}
    \label{Fig2}
\end{figure}

First, both solutions were studied by analytical transmission electron microscopy to characterize their morphology, size, chemical composition, and crystallography. Figure~\ref{Fig2}a shows a high-angle annular dark-field (HAADF) STEM micrograph of prepared nanoparticles. In both solutions, the nanoparticles are mostly spherical; however, other morphologies, such as triangular plates, cubes, squared plates, and nanowires, were obtained in limited numbers as well. Despite continuous growth during synthesis, the nanoparticles are monocrystalline, as demonstrated on a \SI{100}{\nano\meter} nanoparticle. The high-resolution TEM micrograph in Figure~\ref{Fig2}b shows atomic columns in crystallographic orientation [111]. The orientation of the atomic columns is further verified by nano-beam electron diffraction (NBED) from the same area. At the edge of the nanoparticle, an oxide shell is present that is a few nanometer thick. The selected area electron diffractograms (SAED) in Figure~\ref{FigS2} show the predominant crystallographic orientations of the synthesized nanoparticles in both nanoparticle solutions. Crystallographic orientations were identified, using the CrysTBox software \cite{10.1107/S1600576717006793}, as those of a face-centered cubic (FCC) structure of lead \cite{10.1016/S0022-4596-03-00017-3} with Miller indices of (111), (002), (022), (113), (222), (004), (224), and (024). The crystallographic orientations of the nanoparticles in the Pb-NC and Pb-C solutions are identical.

The size distribution is rather wide, as shown in Figure~\ref{Fig2}c, due to the slow cooling of the solution, which allowed nanoparticle growth even after the end of the nucleation period when the temperature of the solution was not sufficiently high to allow further decomposition of lead acetate and formation of nucleation sites in the solution. The nanoparticle size distribution follows log-normal distribution. The Pb-NC solution contains nanoparticles with a mean diameter of \SI{236}{\nano\meter} and a standard deviation of \SI{158}{\nano\meter}, and a mode of \SI{130}{\nano\meter}. The size distribution of the Pb-C solution is less polydisperse, with a mean diameter of \SI{156}{\nano\meter}, a standard deviation of \SI{124}{\nano\meter} and a mode of \SI{63}{\nano\meter}. This difference in the nanoparticle size distribution suggests that, despite the limited solubility in TEG at room temperature, trisodium citrate acts as a weak surfactant at high temperatures, when fully dissolved in the solution. An advantage of trisodium citrate, compared to other surfactants, is the relative ease with which its removal can be carried out \cite{10.1021/acsnano.0c03050}. The washing of the solution successfully removes the residual citrate from the solution, and based on energy dispersive X-ray spectroscopy (EDX) analysis, the citrate adsorbed on the nanoparticle surface is removed by plasma cleaning prior to characterization in TEM (see Figure~\ref{FigS3}). Both Pb-NC and Pb-C samples contain nanoparticles with a few nanometer-thick oxide shell. On the basis of the nearly identical thickness of the oxide shells observed in both solutions, we assume that this minor oxidation is a consequence of the plasma cleaning in the oxygen-argon atmosphere. The identical structural and morphological properties of the nanoparticles from both solutions, along with the identical chemical composition, mean that both solutions can be considered equal in terms of the extrinsic effects on the plasmonic properties of the nanoparticles. 

\begin{figure}[p]
    \centering
    \includegraphics[width=1\linewidth]{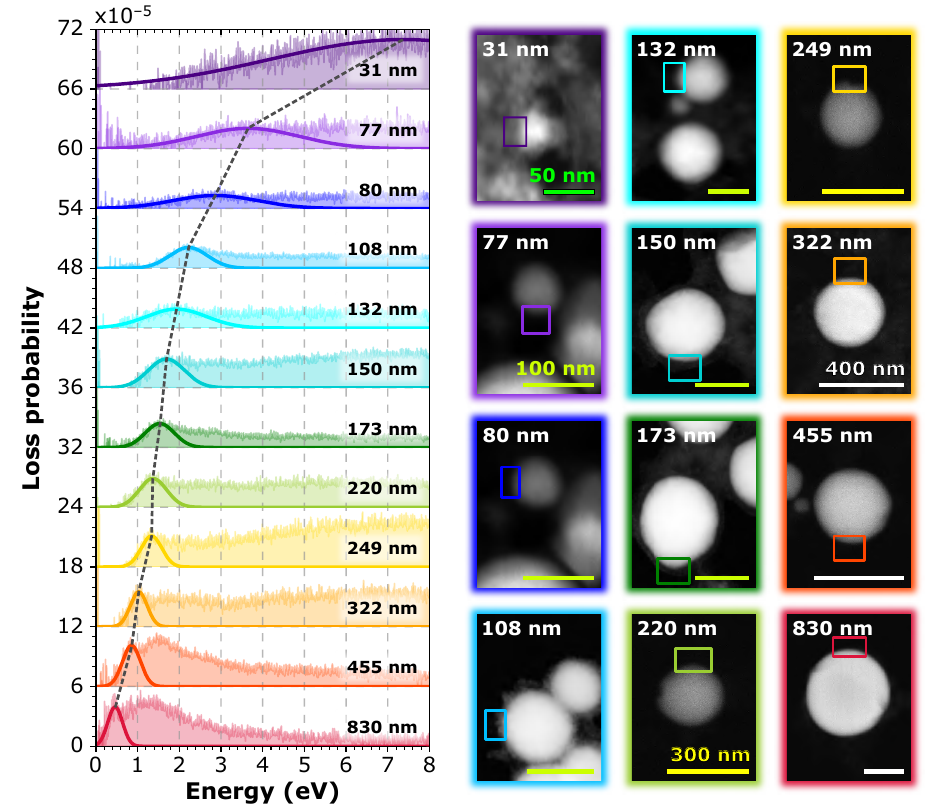}
    \caption{Processed EEL spectra of a set of nanoparticles from the Pb-NC solution with diameters  ranging from 31 to \SI{830}{\nano\meter} with peaks correscponding to the dipole LSPR mode fitted with Gaussians. The rectangles in the ADF STEM micrographs of individual nanoparticles mark the integration areas from which the EEL spectra were collected. The dashed line is intended to serve as a guide for the eye, following the peaks corresponding to the dipole LSPR mode.}
    \label{Fig3}
\end{figure}

Second, we have investigated the plasmonic properties of a set of nanoparticles with diameters ranging from 31 to \SI{830}{\nano\meter} from the Pb-NC solution, as well as nanoparticles from the Pb-C solution  with diameters ranging from 40 to \SI{700}{\nano\meter}. The processed EEL spectra of selected Pb-NC nanoparticles are shown in Figure~\ref{Fig3}. The processed spectra of the selected Pb-C nanoparticles are in Figure~\ref{FigS4}. The colored rectangles in the included ADF STEM micrographs mark the area from which the corresponding EEL spectrum was integrated and collected. The spectra contain pronounced peaks that correspond to the dipolar LSPR peaks. To quantitatively evaluate individual LSPRs, we fitted plasmon peaks with Gaussian functions, giving the plasmon peak energy $E_\mathrm{peak}$, the loss probability maxima $I_\mathrm{max}$, and the full width at half maximum $\Delta E$. The results of the fitting are summarized in Table~\ref{TabS1} for Pb-NC nanoparticles and Table~\ref{TabS2} for Pb-C nanoparticles, respectively.

The loss probability of the dipole LSPR mode is highest for nanoparticles with diameters larger than \SI{300}{\nano\meter}, where the loss probability fluctuates around $3.6\times10^{-5}$. For smaller nanoparticles, the loss probability slowly decreases with decreasing nanoparticle diameter, reaching a minimum of $1.3\times10^{-5}$. The $\Delta E$ of the LSPR dipole mode follows the opposite trend, where it is the lowest for nanoparticles with diameters larger than \SI{200}{\nano\meter}. For nanoparticles with these diameters, $\Delta E$ fluctuates around the value of \SI{0.29}{\electronvolt}. With decreasing nanoparticle diameter, it grows, reaching a value of around \SI{1.25}{\electronvolt} for nanoparticles with a diameter of around \SI{80}{\nano\meter}. Ultimately, it reaches its maximum of \SI{3.36}{\electronvolt} for nanoparticles with diameter around \SI{30}{\nano\meter}. 


\begin{figure}[t]
    \centering
    \includegraphics[width=1\linewidth]{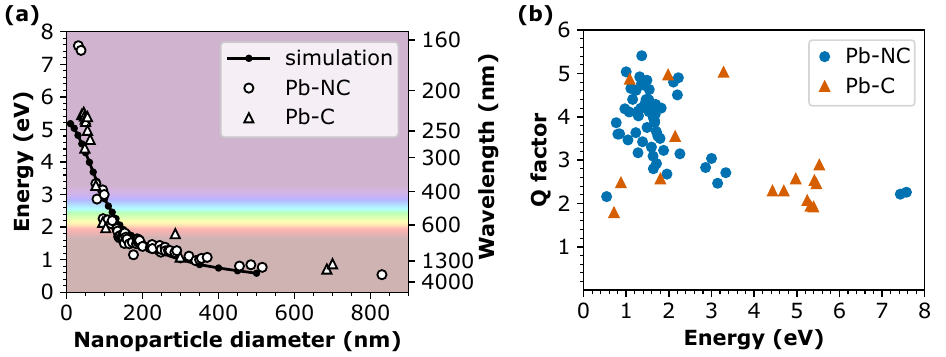}
    \caption{Plasmonic performance of lead nanoparticles. (a) Dipole LSPR energy $E_\mathrm{peak}$ extracted from the measured EEL spectra and from the numerical simulations as a function of the nanoparticle diameter. (b) Experimental Q factors of the dipole LSPR peaks defined as $E_\mathrm{peak}/\Delta E$.}
    \label{Fig4}
\end{figure}

Figure~\ref{Fig4} shows the experimental tunability of the LSPR dipole mode as a function of the diameter of the nanoparticle and the quality factors (Q factors), defined as the peak energy of the LSPR $E_\mathrm{peak}$ divided by its full width at half maximum $\Delta E$. For nanoparticles from the Pb-NC solution, the tunability of the dipole LSPR mod spans from \SI{0.54}{\electronvolt} (corresponding  wavelength \SI{2296}{\nano\meter}) for the \SI{830}{\nano\meter} nanoparticle to \SI{7.57}{\electronvolt} (corresponding wavelength \SI{164}{\nano\meter}) for the \SI{31}{\nano\meter} nanoparticle. The experimental Q factor remains above the value of 2 throughout the entire spectral interval and culminates between the energy of \SI{0.76}{\electronvolt} (wavelength \SI{1631}{\nano\meter}) and 3.34 eV (wavelength \SI{371}{\nano\meter}), achieving an average Q factor of 3.9. The highest observed Q factor reads 5.4 at the LSPR energy of \SI{1.36}{\electronvolt}. We note that the values observed in the Pb-C solution nanoparticles are comparable to those reported above, with minor differences within the experimental error. Compared with other non-noble metals, lead nanoparticles offer higher Q factors in the visible spectral range than bismuth \cite{10.1021/acs.jpclett.5c02531} and gallium \cite{10.1021/acs.jpclett.3c00094, 10.1021/acs.jpclett.5c02035} nanoparticles. In the ultraviolet spectral range, the Q factors of LSPR in lead nanoparticles are comparable to those of gallium nanoparticles \cite{10.1021/acs.jpclett.3c00094, 10.1021/acs.jpclett.5c02035}. 

In addition, we have performed numerical simulations (see Methods) to support our experimental results. The calculated EEL spectra are shown in Figure~\ref{FigS5}. The energy of the dipole mode, extracted from the simulations, is also shown in Figure~\ref{Fig4}a. We generally see a good agreement between the experiment and the numerical simulations while the differences suggest an inaccuracy in the parameters used in the numerical model, including the dielectric function of lead, the approximation of a silicon dioxide membrane by an effective surrounding medium, or the exact shape of the nanoparticles. The main discrepancy in the deep ultraviolet region for nanoparticles smaller than \SI{50}{\nano\meter} is attributed to the dielectric functions of Werner et al. \cite{10.1063/1.3243762} whose real part reaches values below $-2$ up to \SI{6}{\electronvolt} and therefore implies that the energy of the simulated dipole LSPR cannot exceed this value. In contrast, the real part of dielectric functions by Lemonnier et al. \cite{10.1103/physrevb.8.5452} reaches values below $-2$ up to \SI{7}{\electronvolt}. However, the real parts of both two aforementioned dielectric functions of lead are negative up to \SI{12}{\electronvolt}. Consequently, a redefinition of the dielectric function of lead is necessary, particularly in the ultraviolet region, to perform exact numerical simulations. One possible approach would involve the use of spectroscopic ellipsometry or reflectance measurement under ultra-high vacuum conditions, methods that avoid contamination, oxidation, as well as the absorption of air in the deep ultraviolet spectral range.


\begin{figure}[t]
    \centering
    \includegraphics[width=1\linewidth]{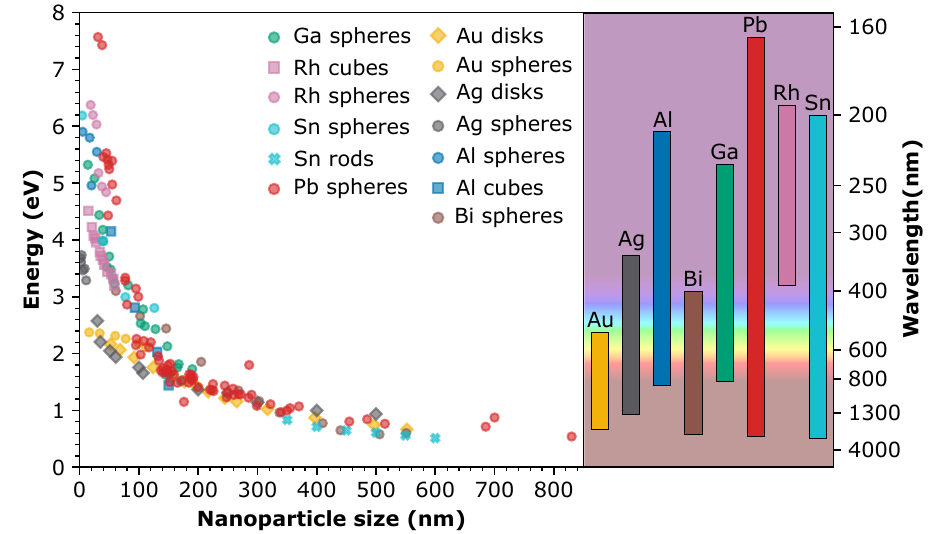}
    \caption{Comparison of spectral tunability of the dipole mode in the lead (Pb) nanoparticles (this work) with other metalic nanostructures including aluminium (Al) \cite{10.1021/acsnano.9b05277, 10.1021/nn405495q, 10.1021/nl303517v, 10.1038/s41598-021-84550-w, 10.1021/nn400918n, 10.1364/ome.3.000954}, gold (Au) \cite{10.1021/nn102166t, 10.1021/ac0702084}, silver (Ag) \cite{10.1021/acsphotonics.7b01060, 10.1021/jp022357q, 10.1038/nature10904}, bismuth (Bi) \cite{10.1021/acs.jpclett.5c02531}, gallium (Ga) \cite{10.1021/acs.jpclett.3c00094, 10.1021/acs.jpclett.5c02035}, rhodium (Rh) \cite{10.1039/c5nh00062a, 10.1039/d4nh00449c}, and tin (Sn) \cite{10.1007/s11468-024-02505-z, 10.1021/acs.jpcc.8b10851}. Note that lead has the widest tunability interval among all presented materials, supporting LSPR resonances from the near-infrared to the deep-ultraviolet spectral region.}
    \label{Fig5}
\end{figure}

Finally, we proceed with a comprehensive comparison of the energy of the dipole LSPR as a function of the nanoparticle size for lead and other metals. To facilitate such a comparison, we compiled the LSPR energies at varying nanoparticle sizes that have been reported in the relevant literature. Given the limited range of nanoparticle diameters that have been examined and documented, a synthesis of the experimental findings from studies of diverse nanoparticle morphologies is warranted. Although the distinct morphology of the nanoparticle results in an imperceptible energy shift of the LSPR, it facilitates the comprehensive assessment of the total LSPR tunability interval of individual metals. This comprehensive comparison is presented in Figure~\ref{Fig5}. We compared the lead (Pb) nanoparticles discussed above with the nanostructures of aluminum (Al) \cite{10.1021/acsnano.9b05277, 10.1021/nn405495q, 10.1021/nl303517v, 10.1038/s41598-021-84550-w, 10.1021/nn400918n, 10.1364/ome.3.000954}, gold (Au) \cite{10.1021/nn102166t, 10.1021/ac0702084}, silver (Ag) \cite{10.1021/acsphotonics.7b01060, 10.1021/jp022357q, 10.1038/nature10904}, bismuth (Bi) \cite{10.1021/acs.jpclett.5c02531}, gallium (Ga) \cite{10.1021/acs.jpclett.3c00094, 10.1021/acs.jpclett.5c02035}, rhodium (Rh) \cite{10.1039/c5nh00062a, 10.1039/d4nh00449c}, and tin (Sn) \cite{10.1007/s11468-024-02505-z, 10.1021/acs.jpcc.8b10851}. Of all the metals, lead has the widest tunability interval, supporting LSPR resonances from the near-infrared to the deep-ultraviolet spectral region, ultimately enabling LSPR generation at energies previously inaccessible. Additionally, there is the possibility of further decreasing the diameter of lead nanoparticles even below \SI{30}{\nano\meter}, reaching LSPR energies above \SI{8}{\electronvolt}. This option is not viable for the other plasmonic metals, where the dimensions reach the physical limit, and thus reaching the LSPR energies above \SI{7}{\electronvolt} is impossible.


\section{Conclusion}

In conclusion, we have synthesized lead nanoparticles through the thermal decomposition of lead acetate trihydrate in tetraethylene glycol. The nanoparticles were characterized by analytical transmission electron microscopy, with the primary focus being placed on STEM-EELS to explore their plasmonic performance and tunability of the localized surface plasmon resonances as a function of the nanoparticle diameter. The findings of this study demonstrate that the tunability of the dipole mode of LSPR encompasses the wavelength range from the near-infrared to the deep-ultraviolet spectral region. This range was identified as the most extensive among all plasmonic elemental metals. Furthermore, the highest LSPR energy observed in lead nanoparticles exceeds those reported for other plasmonic materials used in the ultraviolet spectral region, such as aluminum, rhodium, gallium, and tin. In the deep-ultraviolet spectral region, lead nanoparticles facilitate the generation of LSPR at ultimately high energies above \SI{7}{\electronvolt} that have previously been inaccessible. A comparison of lead nanoparticles with other non-noble metals reveals that they exhibit higher Q factors in the visible spectral range compared with bismuth and gallium nanoparticles. Within the ultraviolet spectral range, the Q factors of LSPR in lead nanoparticles are comparable to those of gallium nanoparticles. Lead nanoparticles have been demonstrated to exhibit stability in ambient air, resulting in the formation of a few nanometer-thick oxide shell, a property that is inherently characteristic of all non-noble metals. The primary disadvantage associated with lead is its toxicity. However, in numerous applications, including metasurfaces, solar cells, and sensors, nanoparticles are affixed to the substrate or can be embedded into a dielectric. In addition, the spectral tunability of the LSPR with the size of lead nanoparticles may serve as a possible linker to optically measure their size by the spectral position of the peak corresponding to the dipole plasmon mode. Consequently, lead nanoparticles offer the widest reported LSPR tunability interval of all plasmonic metals explored to date, thereby demonstrating their potential as a multispectral plasmonic platform.


\section{Methods}

\subsection{Synthesis of nanoparticles}

Monocrystalline lead nanoparticles were synthesized using the thermal decomposition of lead (II) acetate trihydrate. In a typical solution, \SI{30}{\milli\litre} of TEG were heated to \SI{270}{\celsius} in a sand bath. Upon reaching the desired temperature, a solution consisting of \SI{1}{\gram} of lead (II) acetate trihydrate and \SI{15}{\milli\litre} of TEG was injected into the hot TEG and left to react under vigorous stirring for 30 minutes. After the elapsed time was reached, the hot plate was turned off, the vial was removed from the sand bath, and the mixture was left to cool back to room temperature. To obtain solutions with smaller mean nanoparticle diameter and narrower size distribution, an additional \SI{1.5}{\gram} of trisodium citrate (\ce{Na3C6H5O7}) was incorporated into the lead acetate trihydrate-TEG solution prior to injection.

\subsection{Transmission electron microscopy}

The analysis of nanoparticle morphology, size, crystallography, and chemical composition by energy-dispersive X-ray spectroscopy (EDX) was performed on a TEM TFS Talos. All of these TEM measurements performed were performed at \SI{200}{\kilo\electronvolt}.

Plasmonic properties of the nanoparticles were measured by monochromated STEM-EELS at \SI{300}{\kilo\electronvolt} with the convergence semiangle set to \SI{10}{\milli\radian} and the collection semiangle set to \SI{11.4}{\milli\radian}. These parameters were selected to acquire the low-noise EELS signal even for large (and thick) nanoparticles \cite{10.1016/j.ultramic.2020.113044}, while the relativistic background was suppressed by the use of a silicon oxide membrane as a substrate. The Pb-C sample was analyzed using a TEM FEI Titan equipped with a GIF Quantum spectrometer. The probe current was adjusted to around \SI{100}{\pico\ampere}. The dispersion of the spectrometer was set to \SI{0.01}{\electronvolt} per channel, and the full width at half maximum of the zero-loss peak was around \SI{0.18}{\electronvolt}. The acquisition time was adjusted to use the maximal intensity range of the CCD camera in the spectrometer to avoid overexposure. The Pb-NC sample was analyzed using a TEM FEI Titan equipped with a GIF Continuum spectrometer. The probe current was adjusted to around \SI{10}{\pico\ampere}. The dispersion of the spectrometer was set to \SI{0.004}{\electronvolt} per channel, and the full width at half maximum of the zero-loss peak was around \SI{0.12}{\electronvolt}. The acquisition time was adjusted to use the maximal intensity range in the counting mode of the direct detection camera (K3) in the spectrometer to avoid saturation.

EEL spectra were integrated over rectangular areas at the edges of the nanoparticles where the LSPR is significant. They were further divided by the integral intensity of the zero-loss peak to transform the measured counts into a quantity proportional to the loss probability. Next, the EEL spectrum of a pure silicon dioxide membrane was subtracted to remove the background. Finally, the peaks corresponding to the dipole mode of the LSPR were fitted by Gaussian functions reading $$G(E)=I_{\mathrm{max}} \cdot \mathrm{e}^{-4\mathrm{ln}2\cdot\left(\frac{E-E_\mathrm{peak}}{\Delta E}\right)^2}$$ to obtain the peak energy of the plasmon $E_\mathrm{peak}$, the loss probability maxima $I_\mathrm{max}$, and the full width at half maximum $\Delta E$. We note that plasmon peaks have a Lorentzian profile (when considered as damped harmonic oscillators), but EELS measures their convolution with the instrument functions including the energy distribution of the primary electron beam, instrumental instabilities and point spread function of the camera.  Consequently, there is no perfect curve to fit plasmon peaks measured by EELS, and we fit them for simplicity with Gaussians.

\subsection{Numerical simulations}

Numerical simulations of the EEL spectra were performed using the MNPBEM toolbox \cite{10.1016/j.cpc.2015.03.023} based on the boundary element method. Our model consisted of a lead sphere in an effective surrounding medium. The dielectric function of lead by Werner et al. \cite{10.1063/1.3243762} was used and the effective refractive index of the surrounding medium was set to 1.33 to approximate the effect of the silicon dioxide membrane substrate. The \SI{300}{\kilo\electronvolt} electron beam was positioned \SI{5}{\nano\meter} outside the nanoparticle.


\begin{acknowledgement}

This work has been supported by the OP JAK Excellent Research program (project QM4ST, No. CZ.02.01.01/00/22\_008/0004572) and Brno University of Technology (grants No. FSI-S-23-8336). The authors acknowledge the CzechNanoLab Research Infrastructure (ID 90251), supported by MEYS CR. A part of the TEM measurements was realized at Thermo Fisher Scientific in Brno. M.F. acknowledges the Brno Ph.D. Talent Scholarship -- Funded by the Brno City Municipality.

\end{acknowledgement}




\bibliography{literatura}


\newpage
\renewcommand{\thefigure}{S\arabic{figure}}
\renewcommand{\thetable}{S\arabic{table}}
\setcounter{figure}{0}
\setcounter{table}{0}

\begin{figure}[p]
    \centering
    \includegraphics[width=1\linewidth]{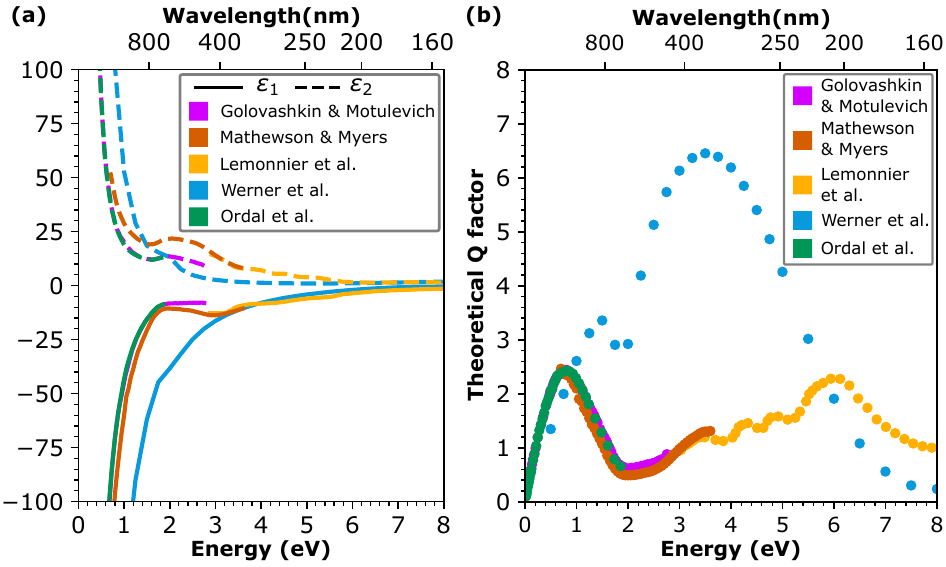}
    \caption{Comparison of measured dielectric functions $\epsilon$ of lead from literature, namely by Werner et al. \cite{10.1063/1.3243762}, Mathewson \& Myers \cite{10.1088/0031-8949/4/6/009}, Ordal et al. \cite{10.1364/ao.26.000744}, Golovashkin \& Motulevich \cite{Golovashkin1968}, and Lemonnier et al. \cite{10.1103/physrevb.8.5452} and theoretical quality factors of localized surface plasmon resonances $Q_\mathrm{LSPR}$ derived out of them as $Q_\mathrm{LSPR}=-\Re{(\epsilon)}/\Im{(\epsilon)}$.}
    \label{FigS1}
\end{figure}

\begin{figure}[p]
    \centering
    \includegraphics[width=1\linewidth]{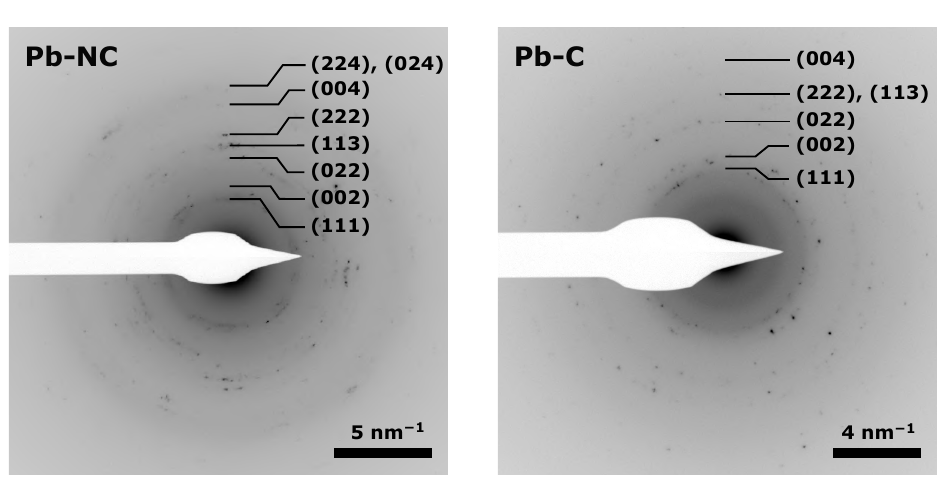}
    \caption{Selected area electron diffractograms (SAED) of synthesized lead nanoparticles in both nanoparticle solutions (Pb-NC and Pb-C). Crystallographic orientations were identified as those of a face-centered cubic structure of lead with Miller indices of (111), (002), (022), (113), (222), (004), (224), and (024). The crystallographic orientations of the nanoparticles in the Pb-NC and Pb-C solutions are identical.}
    \label{FigS2}
\end{figure}

\begin{figure}[p]
    \centering
    \includegraphics[width=0.72\linewidth]{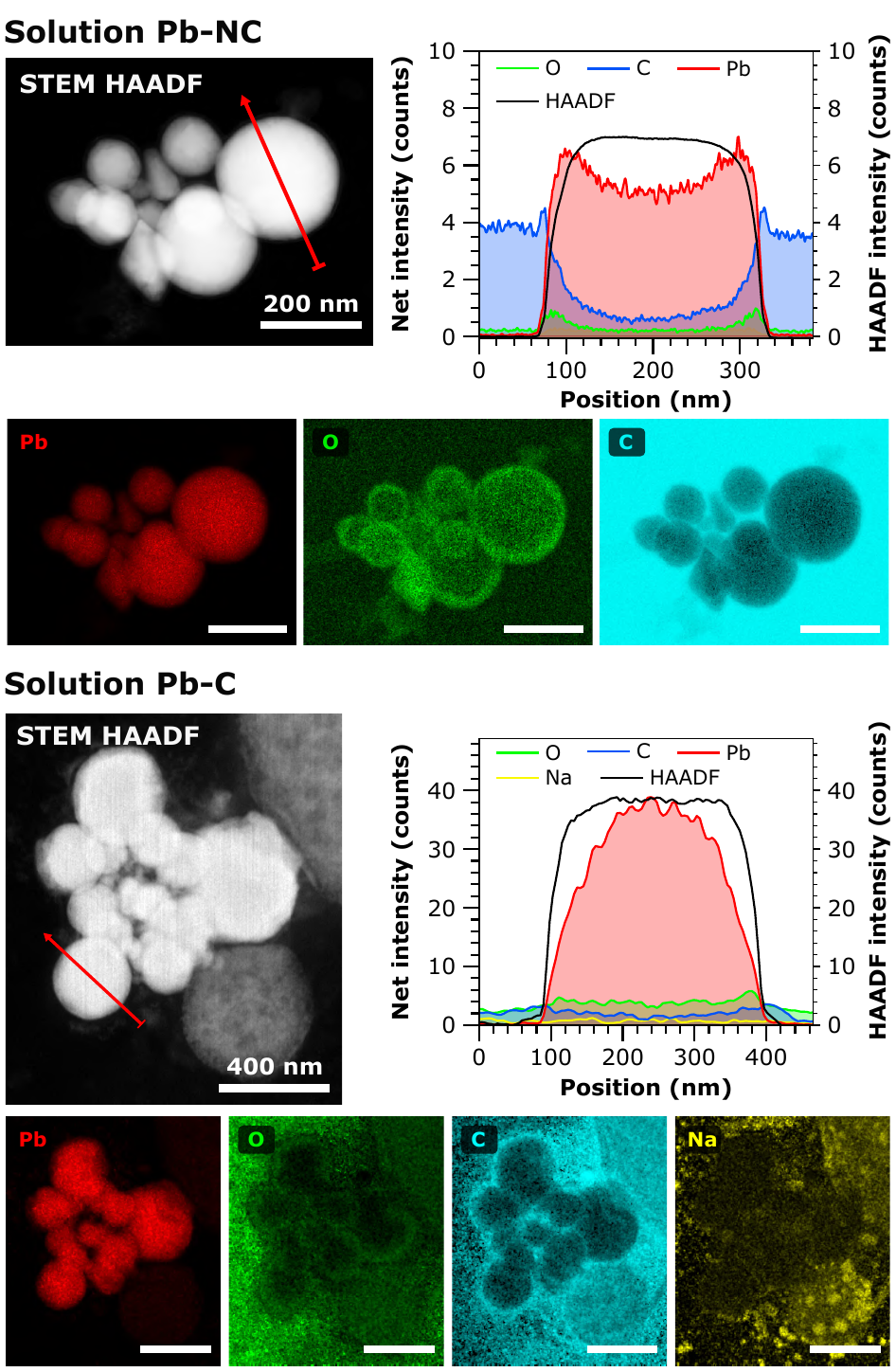}
    \caption{Energy dispersive X-ray spectroscopy (EDX) analysis of synthesized lead nanoparticles in both nanoparticle solutions (Pb-NC and Pb-C) drop casted onto a few-nanometer-thick carbon foil. Elemental maps of lead (Pb), oxygen (O), carbon (C), and sodium (Na, just in the case of Pb-C sample) of nanoparticles from both solutions are shown. In the case of Pb-C sample, the trisodium citrate (\ce{Na3C6H5O7}) adsorbed on the nanoparticle surface is removed by plasma cleaning prior to characterization in TEM while the bulk trisodium citrate remained at some places of the sample as a contamination. Both Pb-NC and Pb-C samples contain nanoparticles with a few nanometer-thick oxide shell.}
    \label{FigS3}
\end{figure}

\begin{figure}[p]
    \centering
    \includegraphics[width=0.9\linewidth]{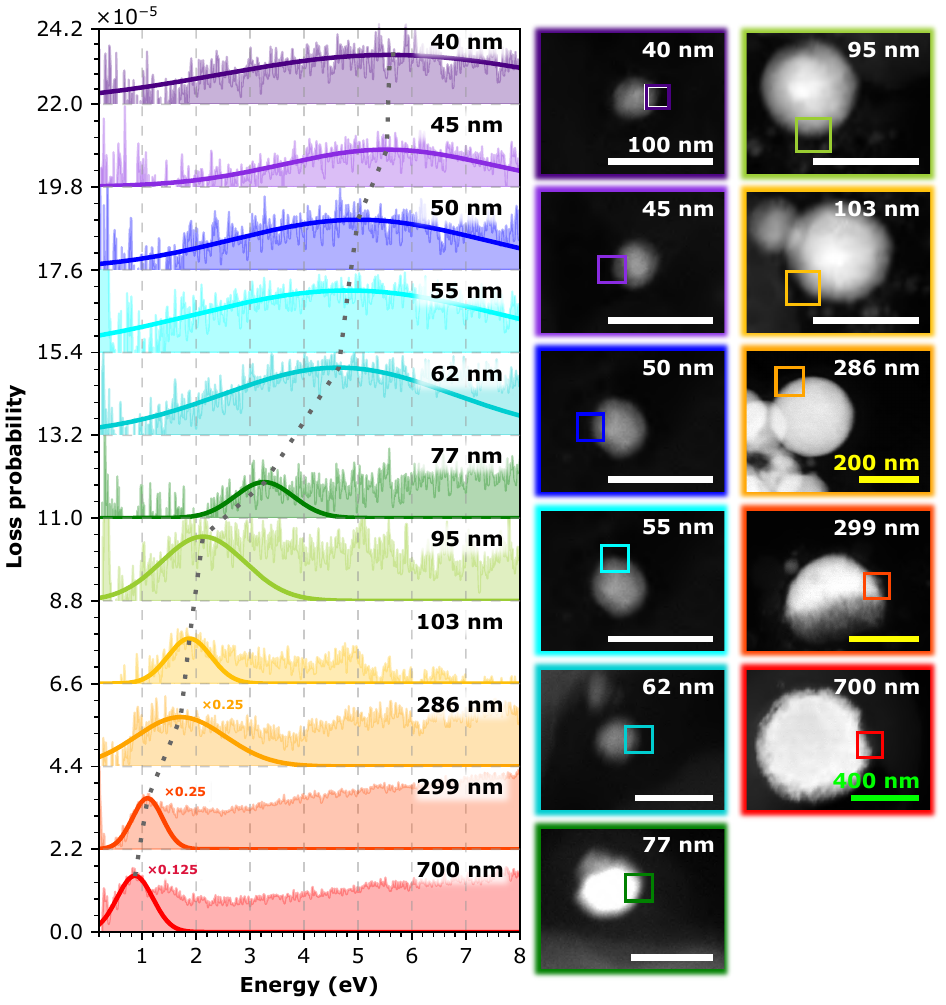}
    \caption{Processed EEL spectra of a set of nanoparticles from solution Pb-C with diameters from 40 to \SI{700}{\nano\meter} with peaks corresponding to the dipole LSPR mode fitted with Gaussians. The rectangles in the ADF STEM micrographs of individual nanoparticles mark the integration areas from which the spectra were collected. The dashed line is intended to serve as a guide for the eye, following the peaks corresponding to the dipole LSPR mode.}
    \label{FigS4}
\end{figure}

\begin{table}[p]
    \centering
    \caption{Quantitative evaluation of individual dipole plasmon peaks in Pb-NC nanoparticles. Plasmon peak energy $E_\mathrm{peak}$, loss probability maxima $I_\mathrm{max}$, and full-width at half-maxima $\Delta E$ are obtained by fitting Gaussian functions. The Q factors are calculated as $E_\mathrm{peak}/\Delta E$.}
    \scriptsize
    \begin{tabular}{c|c|c|c|c}
size (nm) & $E_\mathrm{peak}$ (eV) & $I_{\mathrm{max}}$ & $\Delta E$ (eV) & Q factor \\
\hline
31	&	7.57	& $	4.4 \times 10^{-5}$	&	3.35	&	2.3	\\
38	&	7.43	& $	4.3 \times 10^{-5}$	&	3.36	&	2.2	\\
77	&	3.34	& $	1.8 \times 10^{-5}$	&	1.23	&	2.7	\\
80	&	2.86	& $	1.3 \times 10^{-5}$	&	1.01	&	2.8	\\
95	&	3.14	& $	2.0 \times 10^{-5}$	&	1.27	&	2.5	\\
96	&	2.26	& $	1.6 \times 10^{-5}$	&	0.72	&	3.1	\\
99	&	3.00	& $	2.0 \times 10^{-5}$	&	0.99	&	3.0	\\
108	&	2.22	& $	2.1 \times 10^{-5}$	&	0.45	&	4.9	\\
117	&	2.11	& $	1.7 \times 10^{-5}$	&	0.44	&	4.8	\\
120	&	2.20	& $	2.1 \times 10^{-5}$	&	0.49	&	4.5	\\
132	&	1.96	& $	1.8 \times 10^{-5}$	&	0.73	&	2.7	\\
134	&	1.88	& $	2.9 \times 10^{-5}$	&	0.58	&	3.2	\\
137	&	1.71	& $	1.9 \times 10^{-5}$	&	0.47	&	3.7	\\
140	&	1.67	& $	2.1 \times 10^{-5}$	&	0.43	&	3.9	\\
141	&	1.74	& $	2.2 \times 10^{-5}$	&	0.49	&	3.6	\\
145	&	1.68	& $	1.9 \times 10^{-5}$	&	0.41	&	4.1	\\
148	&	1.71	& $	2.8 \times 10^{-5}$	&	0.59	&	2.9	\\
149	&	1.79	& $	2.6 \times 10^{-5}$	&	0.51	&	3.5	\\
150	&	1.66	& $	2.5 \times 10^{-5}$	&	0.43	&	3.9	\\
152	&	1.50	& $	2.5 \times 10^{-5}$	&	0.34	&	4.4	\\
153	&	1.82	& $	2.1 \times 10^{-5}$	&	0.43	&	4.2	\\
156	&	1.69	& $	2.0 \times 10^{-5}$	&	0.39	&	4.3	\\
159	&	1.63	& $	2.8 \times 10^{-5}$	&	0.41	&	4.0	\\
165	&	1.49	& $	2.3 \times 10^{-5}$	&	0.31	&	4.8	\\
173	&	1.53	& $	2.4 \times 10^{-5}$	&	0.35	&	4.4	\\
176	&	1.15	& $	3.3 \times 10^{-5}$	&	0.26	&	4.4	\\
184	&	1.49	& $	2.8 \times 10^{-5}$	&	0.40	&	3.8	\\
185	&	1.64	& $	2.7 \times 10^{-5}$	&	0.58	&	2.8	\\
187	&	1.58	& $	2.1 \times 10^{-5}$	&	0.37	&	4.3	\\
188	&	1.63	& $	2.7 \times 10^{-5}$	&	0.53	&	3.1	\\
189	&	1.53	& $	2.5 \times 10^{-5}$	&	0.37	&	4.1	\\
191	&	1.54	& $	2.5 \times 10^{-5}$	&	0.33	&	4.6	\\
192	&	1.59	& $	2.2 \times 10^{-5}$	&	0.48	&	3.3	\\
200	&	1.41	& $	2.1 \times 10^{-5}$	&	0.30	&	4.7	\\
220	&	1.36	& $	2.6 \times 10^{-5}$	&	0.25	&	5.4	\\
223	&	1.36	& $	2.4 \times 10^{-5}$	&	0.29	&	4.8	\\
225	&	1.46	& $	2.9 \times 10^{-5}$	&	0.33	&	4.4	\\
226	&	1.34	& $	2.2 \times 10^{-5}$	&	0.32	&	4.2	\\
245	&	1.44	& $	3.1 \times 10^{-5}$	&	0.33	&	4.3	\\
248	&	1.23	& $	1.7 \times 10^{-5}$	&	0.27	&	4.6	\\
249	&	1.32	& $	3.2 \times 10^{-5}$	&	0.27	&	4.9	\\
259	&	1.38	& $	2.9 \times 10^{-5}$	&	0.40	&	3.4	\\
260	&	1.27	& $	2.7 \times 10^{-5}$	&	0.31	&	4.1	\\
268	&	1.29	& $	3.5 \times 10^{-5}$	&	0.32	&	4.0	\\
272	&	1.28	& $	3.2 \times 10^{-5}$	&	0.31	&	4.1	\\
285	&	1.23	& $	2.7 \times 10^{-5}$	&	0.34	&	3.6	\\
290	&	1.28	& $	2.1 \times 10^{-5}$	&	0.40	&	3.2	\\
322	&	1.11	& $	3.4 \times 10^{-5}$	&	0.24	&	4.7	\\
340	&	0.97	& $	3.4 \times 10^{-5}$	&	0.23	&	4.2	\\
350	&	1.00	& $	3.7 \times 10^{-5}$	&	0.20	&	5.0	\\
355	&	1.04	& $	3.7 \times 10^{-5}$	&	0.30	&	3.5	\\
370	&	1.07	& $	3.5 \times 10^{-5}$	&	0.26	&	4.1	\\
455	&	0.80	& $	4.5 \times 10^{-5}$	&	0.22	&	3.6	\\
485	&	0.85	& $	2.8 \times 10^{-5}$	&	0.23	&	3.6	\\
515	&	0.77	& $	3.6 \times 10^{-5}$	&	0.20	&	3.9	\\
830	&	0.54	& $	3.9 \times 10^{-5}$	&	0.25	&	2.2	\\
    \end{tabular}
    \label{TabS1}
\end{table}

\begin{table}[p]
    \centering
    \caption{Quantitative evaluation of individual dipole plasmon peaks in Pb-C nanoparticles. Plasmon peak energy $E_\mathrm{peak}$, loss probability maxima $I_\mathrm{max}$, and full-width at half-maxima $\Delta E$ are obtained by fitting Gaussian functions. The Q factors are calculated as $E_\mathrm{peak}/\Delta E$.}
    \scriptsize
    \begin{tabular}{c|c|c|c|c}
size (nm) & $E_\mathrm{peak}$ (eV) & $I_{\mathrm{max}}$ & $\Delta E$ (eV) & Q factor \\
\hline
40	&	5.46	& $	9.8 \times 10^{-6}$	&	2.22	&	2.5	\\
45	&	5.53	& $	1.2 \times 10^{-5}$	&	1.91	&	2.9	\\
47	&	5.41	& $	9.1 \times 10^{-6}$	&	2.14	&	2.5	\\
48	&	4.43	& $	2.2 \times 10^{-5}$	&	1.94	&	2.3	\\
49	&	5.32	& $	1.3 \times 10^{-5}$	&	2.74	&	1.9	\\
50	&	5.25	& $	1.1 \times 10^{-5}$	&	2.53	&	2.1	\\
55	&	4.98	& $	1.3 \times 10^{-5}$	&	1.94	&	2.6	\\
55	&	5.39	& $	1.5 \times 10^{-5}$	&	2.81	&	1.9	\\
62	&	4.70	& $	2.1 \times 10^{-5}$	&	2.05	&	2.3	\\
77	&	3.28	& $	8.6 \times 10^{-6}$	&	0.65	&	5.0	\\
95	&	2.15	& $	1.5 \times 10^{-5}$	&	0.61	&	3.6	\\
103	&	1.98	& $	1.4 \times 10^{-5}$	&	0.40	&	5.0	\\
286	&	1.80	& $	5.5 \times 10^{-5}$	&	0.70	&	2.6	\\
299	&	1.08	& $	4.8 \times 10^{-5}$	&	0.22	&	4.9	\\
685	&	0.71	& $	8.0 \times 10^{-5}$	&	0.40	&	1.8	\\
700	&	0.88	& $	9.9 \times 10^{-5}$	&	0.35	&	2.5	\\
    \end{tabular}
    \label{TabS2}
\end{table}

\begin{figure}[p]
   \centering
   \includegraphics[width=0.5\linewidth]{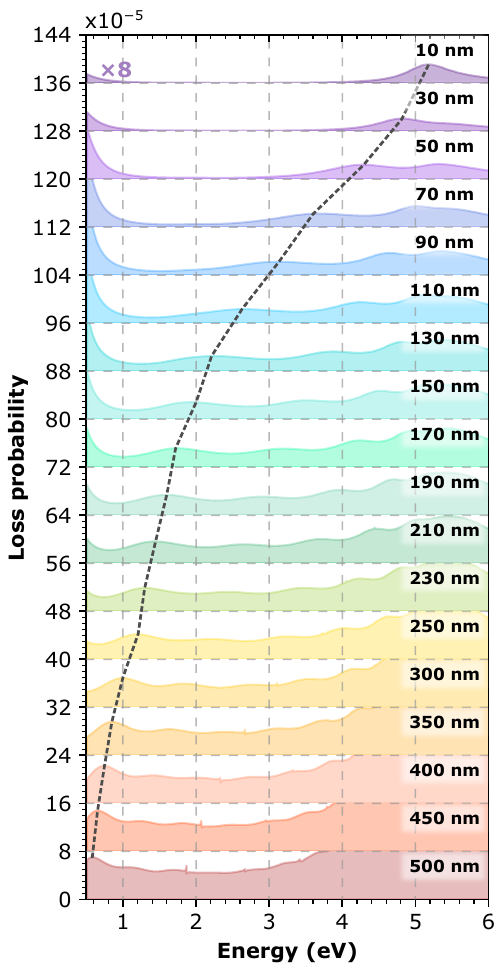}
   \caption{Numerical simulation of EEL spectra of lead nanospheres with the diameters in the range from 10 to \SI{500}{\nano\meter}. The electron beam was situated \SI{5}{\nano\meter} outside the nanosphere. The dashed line is intended to serve as a guide for the eye, following the peaks corresponding to the dipole LSPR mode.}
   \label{FigS5}
\end{figure}

\end{document}